\documentclass[11pt]{article}

\usepackage[margin=1.1in]{geometry}
\usepackage{amsmath,amssymb}
\usepackage{booktabs}
\usepackage{graphicx}
\usepackage{xcolor}
\usepackage{tikz}
\usetikzlibrary{arrows.meta,decorations.pathreplacing}
\usepackage{microtype}
\usepackage[numbers,sort&compress]{natbib}
\usepackage[colorlinks=true,allcolors=blue]{hyperref}

\newcommand{\model}{STRATA\xspace}
\usepackage{xspace}

\title{A Compact Selective State-Space Model for\\
       Cross-Sectional Stock Return Ranking from Raw Intraday Bars}

\author{%
  Mingju Chen\textsuperscript{1} \quad
  Enze Zhang\textsuperscript{1} \quad
  Annan Li\textsuperscript{1} \quad
  Yui Lo\textsuperscript{1}\\
  Kaiming Yu\textsuperscript{1} \quad
  Jinhui Ren\textsuperscript{1} \quad
  Yuanhang Liu\textsuperscript{2,3}\\[0.4em]
  \textsuperscript{1}Famou Agent Team, Baidu AI Cloud\\
  \textsuperscript{2}Qiuzhen College, Tsinghua University\\
  \textsuperscript{3}Institute for AI Industry Research (AIR), Tsinghua University, Beijing, China
}
\date{}

\begin{document}
\maketitle

\begin{abstract}
We present \model (Staggered-Timescale Residual Architecture), a
244{,}633-parameter sequence model that maps five trading days of raw
five-minute bar and order-book data directly to a next-day cross-sectional
return ranking, with no hand-crafted features. The raw-input setting has a
structural obstacle: price series are non-stationary and differ across stocks
by orders of magnitude, so a model easily latches onto price level rather than
dynamics. \model addresses it with a stem of five branches---four learnable
causal depthwise convolutions whose effective kernels are initialised to sum
to zero, plus one cross-field linear contrast---followed by four selective
state-space blocks whose decay biases are staggered across the stack and a
four-path readout. Because a score that merely tilts toward common style
factors scores well on raw rank correlations, every model's scores are
residualised against eight price--volume style factors before any metric is
computed. Trained on four years of data covering roughly one thousand
mid-capitalisation Chinese A-shares and evaluated once on a held-out year,
\model reaches a style-residualised rank information coefficient of $0.0728$
(information ratio $1.128$, signal long-short Sharpe $12.85$), ahead of six
parameter-matched sequence baselines on all four reported metrics; on rank IC
the day-level paired gap against every baseline is significant at $p<0.001$,
and among the arms competitive on predictive power \model's scores are the
least explained by the controls. The close-to-close target opens before the
score exists: measured instead from the first executable price, the decile
spread is indistinguishable from zero, while the ordering of the seven
architectures is unchanged and \model's margin widens.
\end{abstract}

\section{Introduction}

Ranking a cross-section of stocks by their next-day return is the elementary
prediction problem behind systematic equity investing. On any trading day $t$ a
model observes recent market data for each of $N_t$ stocks and must emit one
score per stock; only the \emph{ordering} of those scores matters, because
positions are formed by sorting.

The dominant practice is to solve this problem in two stages. A library of
hand-designed features---formulaic alphas~\citep{kakushadze2016alphas},
rolling technical statistics, and fundamental ratios---is computed first, and a
learned model consumes that feature table~\citep{yang2020qlib}. The
two-stage design is a strong engineering default, but it fixes in advance which
functions of the raw data the model may see. A model reading raw bars is free to
form its own; the cost is that it must solve, internally, the problems that
feature engineering solves externally.

Two such problems dominate. The first is \textbf{scale and level
heterogeneity}. Prices differ across stocks by orders of magnitude and drift
within a stock over years, so the raw input is dominated by a component---the
price level---that is nearly constant within a sample and almost uninformative
about next-day returns. Hand-built features avoid this by construction: returns,
ratios and $z$-scores are all level-free. The second is \textbf{style
confounding}. A score that is a monotone function of, say, market
capitalisation or realised volatility will exhibit a substantial cross-sectional
rank correlation with returns during any period in which that style is
rewarded, without containing information specific to individual stocks. Raw
correlations may therefore reflect rewarded style exposures rather than
stock-level predictive information.

This paper describes \model, a model designed for the raw-input setting under a
deployment-realistic constraint set, and evaluates it under a protocol that
removes exposure to eight price--volume style factors before any metric is
computed. Timescales are staggered in two places, across the three
receptive fields of the stem and across the decay biases of the state-space
stack, and four of the stem's five branches are residuals---deviations of a
channel from its own recent history---rather than the channel itself.
The experiments impose three input constraints: (i) every input channel is a raw field
supplied by the data source---no cross-field algebra, no rolling statistics, no
factor synthesis, no dimensionality reduction; (ii) input preprocessing is
restricted to per-field constant rescaling, a per-field logarithmic transform,
per-field standardisation, and missing-value filling; and (iii) every
preprocessing statistic is estimated on the training split alone.

Our contributions are:

\begin{itemize}
\item \textbf{An architecture for level-robust representation of raw bars.} The
  stem of \model replaces hand-built differencing and rolling statistics with
  four fully trainable causal depthwise convolution branches, initialised so
  that each branch's effective kernel sums to zero, alongside a single
  cross-field linear contrast that mixes fields within a time step.
  Zero-sum-constrained
  convolution as a learnable high-pass front-end is due to
  \citet{bayar2016constrained,bayar2018tifs}; our variant is causal, applied
  along time in one dimension, and paired with learnable smoothing kernels at
  three receptive fields, so that a single stem produces deviations from the
  channel's own recent history at several horizons (Section~\ref{sec:stem}).
\item \textbf{A compact selective state-space backbone with staggered
  timescales.} Four selective state-space blocks are stacked with decay biases
  spread linearly across the stack, and the sequence recursion is evaluated by a
  log-depth associative scan~\citep{hillis1986dataparallel} rather than a time
  loop. A four-path readout summarises the sequence at several effective
  horizons (Sections~\ref{sec:ssm} and~\ref{sec:readout}).
\item \textbf{A controlled comparison under style-residualised evaluation.}
  \model is compared against six standard sequence architectures matched to
  within $\pm5\%$ of its parameter count, sharing the same data, loss,
  optimiser, epoch budget and evaluator. All reported metrics are computed
  after residualising scores against eight style factors, and two robustness
  analyses accompany the comparison: a style-exposure analysis showing that
  the lead is not explained by greater exposure to the controls, and a timing
  decomposition separating the prediction benchmark from the executable
  return (Section~\ref{sec:exp}).
\end{itemize}

\section{Related Work}

\paragraph{Cross-sectional return prediction.}
Empirical asset pricing organises the cross-section of returns around a small
number of factors~\citep{fama1993factors}, with momentum among the most durable
of the anomalies studied since~\citet{jegadeesh1993momentum}. In practice,
quantitative signals are frequently expressed as libraries of formulaic
expressions over price and volume~\citep{kakushadze2016alphas}, and open
platforms distribute feature tables assembled in this
style~\citep{yang2020qlib}. Learned factor models such as
FactorVAE~\citep{duan2022factorvae} keep the factor structure but replace
hand-specified loadings with a latent variable model. All of these consume
engineered inputs, whereas \model consumes raw bars and is therefore compared
with alternative sequence architectures. We use the standard active-management
metrics of information coefficient and information ratio~\citep{grinold1989fundamental}.

\paragraph{Sequence architectures.}
Recurrent networks~\citep{hochreiter1997lstm,cho2014gru}, dilated causal
convolutions~\citep{bai2018tcn} and self-attention~\citep{vaswani2017attention}
are the standard encoders for time series, and patch-based transformers such as
PatchTST~\citep{nie2023patchtst} adapt attention to long horizons; simple linear
maps remain a strong baseline for forecasting~\citep{zeng2023dlinear}.
Structured state-space models~\citep{gu2022s4} and their selective variant,
Mamba~\citep{gu2023mamba}, obtain long effective memory at linear cost in
sequence length, and have been applied to financial series~\citep{mehrabian2025samba}.
Unlike forecasting, the present task maps each input sequence to a scalar that
must be comparable across stocks on the same day.

\paragraph{Learnable decomposition front-ends.}
Constraining the first convolutional layer to zero-sum kernels---forcing a
high-pass, prediction-residual filter---was introduced for image forensics
by~\citet{bayar2016constrained,bayar2018tifs}, following a longer line of
hand-designed residual filter banks in steganalysis~\citep{fridrich2012rich};
the motivation transfers directly to raw price sequences, where the ``direct
current'' component is the price level. Decomposing a series into a smoothed
trend and a residual is likewise a recurring theme in forecasting:
Autoformer~\citep{wu2021autoformer} and FEDformer~\citep{zhou2022fedformer}
build moving-average blocks into the encoder, and
DLinear~\citep{zeng2023dlinear} shows that a simple decomposition followed by
a linear map is a strong model. The stem of \model sits at the meeting point
of the two lines: a causal, depthwise, fully learnable front-end whose
zero-sum-initialised kernels at three receptive fields produce deviations
from the channel's own recent history at several horizons at once.
RevIN~\citep{kim2022revin} attacks the level problem from the normalisation
side with per-instance statistics, which the preprocessing constraint of
Section~\ref{sec:preproc} excludes.

\section{Model}
\label{sec:model}

\begin{figure}[t]
  \centering
  \includegraphics[width=\linewidth]{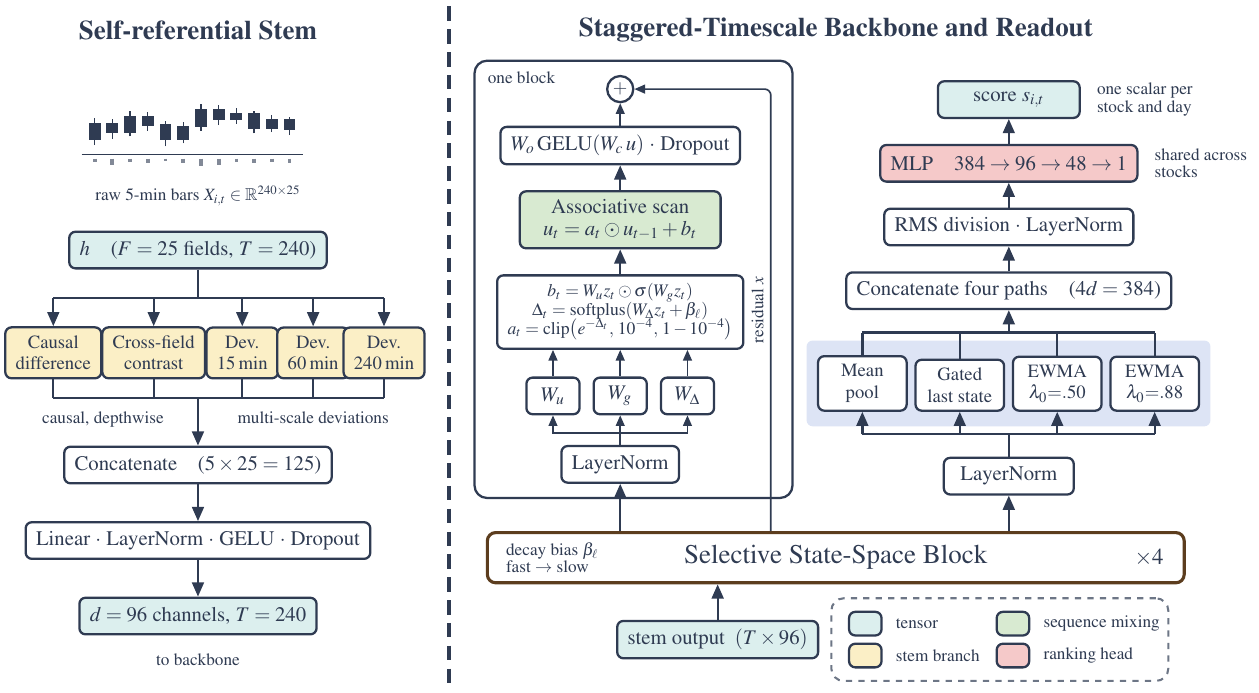}
  \caption{Overview of \model.  Raw five-minute bars are transformed by a
  self-referential stem (trainable causal differences, a cross-field contrast,
  and three multi-scale deviations), processed by four selective state-space
  blocks with staggered decay initialisations, and summarised by four temporal
  paths before per-stock normalisation and the shared MLP ranking head.  The
  tensor widths shown in the diagram correspond to the default configuration
  ($F=25$, $T=240$, $d=96$).}
  \label{fig:architecture}
\end{figure}

Figure~\ref{fig:architecture} provides a compact overview of the complete
forward path.  Each stock is first represented by five days of raw
five-minute fields; the self-referential stem then forms level-robust,
multi-scale residual channels before the selective state-space backbone mixes
information over staggered timescales.  A four-path readout, followed by
per-stock normalisation and the shared ranking head, produces the scalar score
used for cross-sectional ordering.  The following subsections define these
stages precisely and state the temporal and preprocessing conventions that
make the diagram causal.

\subsection{Problem setup}
\label{sec:setup}

\paragraph{Temporal protocol.} Fix a stock $i$ and a \emph{signal day} $t$. Let
\begin{equation}
X_{i,t}\;=\;\big[\,b_{i,t-D+1},\,\ldots,\,b_{i,t-1},\,b_{i,t}\,\big]
\;\in\;\mathbb{R}^{T\times F}
\label{eq:window}
\end{equation}
where $b_{i,d}\in\mathbb{R}^{B\times F}$ are the $B$ intraday bars of day $d$,
so that $T=D\cdot B$. In all experiments $D=5$, $B=48$ (five-minute bars),
$T=240$ and $F=25$. Three properties of Equation~\eqref{eq:window} are worth
naming:

\begin{itemize}
\item The window is $t-4,\,t-3,\,t-2,\,t-1,\,t$: \textbf{five complete trading
      days ending on, and including, the signal day itself}, not the five
      days before $t$.
\item The last element of the window is the final five-minute bar of day $t$,
      which covers the closing minutes of that session. The score
      $s_{i,t}=f_\theta(X_{i,t})$ is therefore produced only \emph{after the
      final five-minute bar of the signal day becomes available}.
\item The learning and evaluation target is the next-day adjusted close-to-close
      return
      \begin{equation}
      y_{i,t}\;=\;\frac{C^{\mathrm{adj}}_{i,t+1}}{C^{\mathrm{adj}}_{i,t}}-1 ,
      \label{eq:label}
      \end{equation}
      whose starting price $C^{\mathrm{adj}}_{i,t}$ is the close of the last bar
      already contained in $X_{i,t}$. The superscript marks the only place in
      the problem where corporate-action adjustment appears; the inputs
      themselves are unadjusted, and the convention is stated in full in
      Section~\ref{sec:preproc}.
\end{itemize}

The quantity of interest is the cross-sectional Spearman correlation between
$\{s_{i,t}\}_i$ and $\{y_{i,t}\}_i$ on each day $t$.

\begin{figure}[t]
\centering
\includegraphics[width=.92\linewidth]{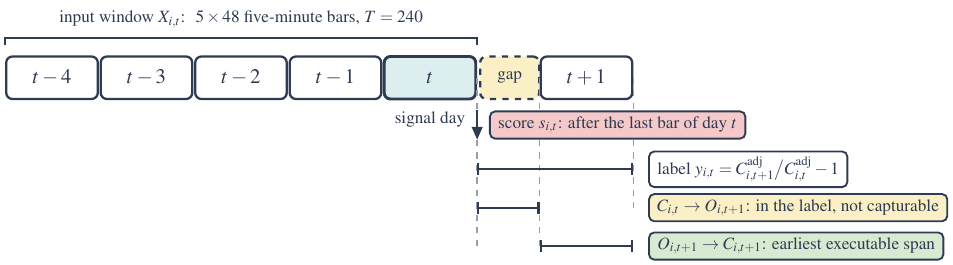}
\caption{Temporal protocol. The input window \emph{includes} the signal day
$t$ and ends with its final five-minute bar; the score is produced only after
that bar is available. The label spans $t\rightarrow t+1$ measured
close-to-close, so its first segment---the overnight move from the day-$t$
close to the day-$t{+}1$ open---lies before the earliest moment at which the
score could be traded. No input price postdates the score, so there is no
look-ahead bias; but the label is not an executable return.}
\label{fig:timeline}
\end{figure}

\paragraph{No look-ahead bias is not the same as a directly tradable backtest.}
Equation~\eqref{eq:label} introduces no look-ahead information: every price in
$X_{i,t}$ precedes the start of the target span, and $C^{\mathrm{adj}}_{i,t+1}$
enters only as a label, never as an input. But the score exists only after the
day-$t$ close at which the target opens, so the overnight segment
$C_{i,t}\rightarrow O_{i,t+1}$ inside the label cannot be captured by any
strategy acting on the score; the close-to-close target is a \emph{prediction
benchmark}, not an exactly replicable trading return.
Figure~\ref{fig:timeline} summarises the ordering; Section~\ref{sec:exec}
measures how much of the reported performance that segment accounts for, and
Section~\ref{sec:limits} states the consequences.

\subsection{Input preprocessing}
\label{sec:preproc}

Preprocessing is deliberately minimal and identical at training and inference
time. Each field is multiplied by a fixed per-field constant that fixes its
unit; price and size fields then receive a $\log(1+x)$ transform; every field is
standardised by its own mean and standard deviation; remaining missing entries
are set to zero after standardisation. The means and standard deviations are
accumulated over the training split only and stored with the weights.
No statistic used here is computed per sample or per stock, and no channel is a
function of more than one raw field.

\paragraph{Adjustment convention.} Inputs are unadjusted: every channel enters
the network exactly as printed by the exchange, and the adjustment factor is
not among the $F=25$ fields. Corporate-action adjustment enters only through
the label of Equation~\eqref{eq:label}, whose adjusted closes multiply the raw
close by the day's cumulative backward adjustment factor---a quantity
determined by actions with ex-dates on or before that day and never revised by
later ones. Two properties follow: no corporate-action information can reach
an input retroactively, and an ex-date inside the five-day window leaves its
mechanical gap in the raw series for the model to absorb.
Appendix~\ref{app:adjust} states the convention in full and derives both.

For price fields large relative to one, the logarithmic transform followed by
the initial difference kernel approximates a log return. The kernel remains
trainable, and the cross-field contrast branch reads the standardised channel
values directly, retaining level information alongside the four residual
branches.

\subsection{Self-referential stem}
\label{sec:stem}

Write $h\in\mathbb{R}^{B\times F\times T}$ for the preprocessed input with
channels first. The stem forms $2+S$ branches, all of them depthwise
(\texttt{groups}$=F$: one kernel per field, no field mixing) except one, and
concatenates them along the channel axis.

\paragraph{Causal difference.} A depthwise kernel $k^{\mathrm{d}}\in\mathbb{R}^{F\times 1\times 2}$
with left padding of one step,
\begin{equation}
  \mathrm{diff}_{f,t} \;=\; k^{\mathrm{d}}_{f,0}\,h_{f,t-1} + k^{\mathrm{d}}_{f,1}\,h_{f,t},
  \qquad k^{\mathrm{d}}_{f,\cdot}\;\text{initialised to}\;(-1,+1).
\end{equation}
The kernel is a free parameter; only its initialisation is the first difference.

\paragraph{Cross-field contrast.} A single bias-free linear map
$W^{\mathrm{c}}\in\mathbb{R}^{F\times F}$ applied at each time step,
$\mathrm{con}_{t}=W^{\mathrm{c}}x_t$. This is the one branch that mixes fields,
and it mixes them only within a time step; which combinations are useful is left
to the gradient.

\paragraph{Multi-scale deviation from self.} For each scale $w\in\mathcal{S}$
a depthwise kernel $k^{(w)}\in\mathbb{R}^{F\times 1\times w}$, initialised
uniformly to $1/w$, is convolved causally with $h$, and the branch output is
the residual, gated by a per-scale per-field positive gain
$\gamma^{(w)}_f=\mathrm{softplus}(\rho^{(w)}_f)$ with $\rho$ initialised to zero:
\begin{equation}
  \mathrm{res}^{(w)}_{f,t} \;=\; \gamma^{(w)}_f\Bigl(h_{f,t} - \textstyle\sum_{u=0}^{w-1}k^{(w)}_{f,u}\,h_{f,\,t-w+1+u}\Bigr).
  \label{eq:res}
\end{equation}
Equation~\eqref{eq:res} is algebraically a single depthwise causal convolution
with kernel $\delta_{\text{last}}-k^{(w)}$, every tap of which is learned. At
initialisation that kernel sums to zero, which is exactly the constrained
high-pass form of~\citet{bayar2016constrained}; unlike that work we do not
re-impose the constraint after each update, so the branch may drift away from
being strictly zero-sum if the data prefer it. We use
$\mathcal{S}=\{3,12,48\}$ bars, i.e.\ $15$, $60$ and $240$ minutes, the last of
which is one trading session.

The $F(2+S)$ concatenated channels are projected to width $d$ by
$\mathrm{Linear}\rightarrow\mathrm{LayerNorm}\rightarrow\mathrm{GELU}\rightarrow\mathrm{Dropout}$.
With $F=25$, $S=3$ and $d=96$ the projection reads $125\rightarrow96$.

\subsection{Selective state-space blocks}
\label{sec:ssm}

The backbone is $L$ residual blocks. Block $\ell$ maps
$x\in\mathbb{R}^{B\times T\times d}$ to
\begin{align}
  z &= \mathrm{LayerNorm}(x), \qquad
  b = (W_u z)\odot\sigma(W_g z), \\
  \Delta &= \mathrm{softplus}(W_\Delta z + \beta_\ell), \qquad
  a = \mathrm{clip}\bigl(e^{-\Delta},\,10^{-4},\,0.9999\bigr),
  \label{eq:gates}
\end{align}
followed by the first-order recursion and the residual output
\begin{equation}
  u_t = a_t\odot u_{t-1} + b_t,
  \qquad
  \mathrm{Block}_\ell(x) = x + \mathrm{Dropout}\bigl(W_o\,\mathrm{GELU}(W_c u)\bigr).
\end{equation}
Both the decay $a_t$ and the drive $b_t$ depend on the input at time $t$,
which makes the block selective in the sense
of~\citet{gu2023mamba}.

The recursion is evaluated with a Hillis--Steele associative
scan~\citep{hillis1986dataparallel} over the pair $(a,b)$ under the composition
$(a_1,b_1)\circ(a_2,b_2)=(a_1a_2,\;a_2b_1+b_2)$, giving $\lceil\log_2 T\rceil$
sequential steps instead of $T$. For $T=240$ this is $8$ passes; the total work
is $\mathcal{O}(T\log T)$ and the critical-path depth is $\mathcal{O}(\log T)$.

The per-channel bias $\beta_\ell$ is the only quantity that differs across
blocks: it is initialised to $L$ values spread linearly from $1.0$ down to
$-1.5$, so the stack begins with fast-forgetting blocks at the bottom and
slow-forgetting blocks at the top. With $L=4$ the initial half-lives
$\ln 2/\Delta$ are $0.53$, $0.89$, $1.67$ and $3.44$ bars---between roughly two
and seventeen minutes, a $6.5\times$ spread. Because $a$ is clipped below
$1$ rather than fixed, the trained half-lives are free to move far beyond
their initial values.

\subsection{Readout}
\label{sec:readout}

Let $u\in\mathbb{R}^{B\times T\times d}$ be the final block output after a
LayerNorm. Four summaries are concatenated: the mean over time; the last step,
scaled by a learnable gate $\sigma(g)$ with $g$ initialised to $-1$ so that the
model does not begin by relying on a single bar; and two exponentially weighted
averages $\sum_t w_t u_t$ with $w_t\propto \lambda^{\,T-1-t}$ for two learnable
decays $\lambda=\sigma(\cdot)$ initialised to $0.5$ and $0.88$. The concatenated
vector $v\in\mathbb{R}^{4d}$ is then divided by its own root-mean-square,
\begin{equation}
  \tilde v \;=\; v \big/ \sqrt{\textstyle\frac{1}{4d}\sum_j v_j^2 + \epsilon},
  \label{eq:rms}
\end{equation}
passed through a LayerNorm, and mapped to a scalar by a three-layer MLP
($4d\rightarrow d\rightarrow d/2\rightarrow 1$).

The division in Equation~\eqref{eq:rms} is redundant given the LayerNorm that
follows it.\footnote{LayerNorm is invariant to positive rescaling,
$\mathrm{LN}(cv)=\mathrm{LN}(v)$ for $c>0$; the division is retained only to
document the trained implementation.}

\subsection{Objective}
\label{sec:loss}

Batches are formed by trading day, so a batch is a cross-section and a
within-batch correlation is an estimate of the daily information coefficient.
Labels are clipped to the $[0.75,\,99.25]$ percentile range of the training
split and then standardised within each day. Writing $p$ for predictions and
$y$ for labels within a batch of size $n$, and $c$ for a weight ramped
along a cosine schedule from $0.03$ to $0.30$ over the first epochs,
\begin{equation}
  \mathcal{L}_{\mathrm{opt}} = \tfrac{1}{2}\,\mathrm{Huber}_\beta(p,y)
  \;-\;(0.5+c)\,\mathrm{corr}\bigl(\tilde{r}(p),\tilde{r}(y)\bigr)
  \;-\;c\,\mathrm{corr}(p,y)
  \;+0.02\,\bigl[\overline{\hat p^4}-3\bigr]_+ .
  \label{eq:loss}
\end{equation}
where $\tilde r$ is a differentiable rank obtained from pairwise sigmoids,
$\hat p$ is the standardised prediction, and $[\cdot]_+$ is the positive
part. The four active terms are a point-prediction anchor, a soft Spearman
surrogate, a Pearson term and a one-sided excess-kurtosis penalty. The pairwise
rank is computed on a subsample of at most $1300$ names per batch to bound its
$\mathcal{O}(n^2)$ cost, and only during training.

For monitoring we additionally report the label spread between the predicted
top and bottom deciles,
$D_{\mathrm{top-bottom}}=(\bar y_{\mathrm{top}}-\bar y_{\mathrm{bot}})/\mathrm{std}(y)$,
where the bars denote labels selected by an $\texttt{argsort}$ of $p$.
This selection is non-differentiable, so $D_{\mathrm{top-bottom}}$ is a logged
diagnostic and is not part of the optimised objective.

\subsection{Parameter budget}

Table~\ref{tab:params} breaks down the $244{,}633$ trainable parameters.

\begin{table}[t]
\centering
\caption{Parameter breakdown of \model at $d=96$, $L=4$, $\mathcal{S}=\{3,12,48\}$,
$F=25$.}
\label{tab:params}
\begin{tabular}{lrr}
\toprule
Component & Parameters & Share \\
\midrule
Self-referential stem            & 14{,}613  & 5.97\% \\
Selective SSM blocks ($\times4$) & 187{,}392 & 76.60\% \\
LayerNorms (final, feature)      & 960       & 0.39\% \\
MLP head                         & 41{,}665  & 17.03\% \\
Readout scalars ($g$, two decays)& 3         & $<$0.01\% \\
\midrule
Total                            & 244{,}633 & 100\% \\
\bottomrule
\end{tabular}
\end{table}

\section{Experiments}
\label{sec:exp}

\subsection{Data}
\label{sec:data}

We use five-minute bars for a universe of mid-capitalisation Chinese A-shares---the historical constituents of the CSI~1000 index---covering 2019-01-02 through 2024-12-31, $1{,}456$ trading days. The universe is
defined by the data source on each day and contains close to one thousand names
per day ($995$--$999$ median across splits, minimum $878$); membership is
point-in-time, so a stock enters and leaves the panel on the dates it actually
entered and left the index, and no survivorship filter is applied.
Each stock-day is exactly $48$ bars. The $F=25$ fields per bar are open, high,
low, close, volume, turnover, trade count, and three levels of ask and bid
price, volume and order count. Labels are the next-day adjusted close-to-close
returns of Equation~\eqref{eq:label}; inputs remain unadjusted, as specified in
Section~\ref{sec:preproc} and Appendix~\ref{app:adjust}.

\paragraph{Splits.} The nominal training window is 2019-01-02 to 2022-12-30
($972$ trading days). The validation year 2023 ($242$ days, $240{,}204$ samples)
is used for model and protocol selection; the 2024 test year ($241$ evaluable
days) is not used for training or checkpoint selection. All post-hoc
analyses reuse the stored test-year scores. The training window is truncated so that no
training label uses a price observed on or after the first validation day: with
a one-day horizon and a two-day embargo, the last training day is 2022-12-27,
three trading days before validation begins, leaving $955{,}061$ training
samples.

\subsection{Evaluation protocol}
\label{sec:protocol}

Raw model scores are never scored directly. On each trading day the scores of
that day's cross-section pass through the following pipeline before any metric
is computed:

\begin{enumerate}
\item non-finite scores are dropped;
\item scores are winsorised to the cross-sectional mean $\pm3$ standard
      deviations, and the eight style regressors of step~4 are winsorised by
      the same rule;
\item scores are standardised cross-sectionally;
\item scores are regressed by ordinary least squares on eight style factors
      and \textbf{the residual replaces the score}.
\end{enumerate}

The eight styles are computed from the same bars, causally, and are
deliberately simple. \textsc{size} is the $\log$ of $20$-day mean turnover
(a proxy for market capitalisation; direct float data are not available in
this dataset), \textsc{sizenl} is the cube of the cross-sectional $z$-score of \textsc{size},
\textsc{liquidity} ($\log$ $5$-day mean turnover minus $\log$ $60$-day mean),
\textsc{resvol} ($20$-day standard deviation of past log returns),
\textsc{momentum} (the sum of past log returns over the $120$-day window ending
at $t$ minus the sum over the most recent $20$ days, i.e.\ a $100$-day sum
ending $20$ days back),
\textsc{beta} (rolling $60$-day regression slope on the equal-weight
market return), \textsc{strev} ($5$-day sum of past log returns), and
\textsc{intravol} (the mean, over the day's five-minute bars, of each bar's
high-low range divided by its close, then averaged over $20$ days).
Coverage is $93.1\%$--$98.2\%$ per style on the test year.

Several controls have non-zero rank IC on this universe, so residualisation
removes variance that predicts returns. Residualised ICs reported here are
therefore not directly comparable with raw ICs reported elsewhere.

Scores and regressors are clipped symmetrically. Positive-control checks in
Appendix~\ref{app:controls} shows that clipping only the score leaves a residual
tail artefact; all residualised results use the symmetric protocol.

Four metrics are reported on the residualised scores:
\textbf{Rank IC}, the mean over days of the cross-sectional Spearman
correlation with the labels of Equation~\eqref{eq:label};
\textbf{IC\_IR}, that mean divided by its standard deviation across days;
\textbf{Signal LS Sharpe}, the annualised Sharpe ratio obtained by applying
equal-weight long and short decile portfolios, formed on the residualised
score, to the same next-day close-to-close returns that serve as prediction
labels; and
\textbf{Stress IC\_IR}, the IC\_IR restricted to stress days. Signal LS Sharpe
is a frictionless diagnostic against the prediction target, not an executable
strategy return (Section~\ref{sec:exec}). Stress days are selected by a
validation-stage rule based only on market returns; its full specification and
non-causal partitioning convention are given in Appendix~\ref{app:stress}. The
test year contains $55$ stress days in four segments.

\subsection{Baselines}
\label{sec:baselines}

Six standard encoders are compared, each wrapped in the same mean-and-last
pooling head rather than the four-path readout of Section~\ref{sec:readout}.
\textbf{MLP} flattens the whole $T\times F$ window; \textbf{LSTM} and
\textbf{GRU} are two-layer recurrent networks over the raw field vector at each
step~\citep{hochreiter1997lstm,cho2014gru}; \textbf{TCN} is a stack of four
dilated causal convolution blocks~\citep{bai2018tcn}; \textbf{Transformer} is a
two-layer pre-norm encoder with learned positional
embeddings~\citep{vaswani2017attention}; and \textbf{Mamba} is a selective
state-space stack with a linear stem, a single decay initialisation and
mean-and-last pooling, i.e.\ the backbone of \model with each of the components
of Section~\ref{sec:model} removed~\citep{gu2023mamba}.

Each baseline's hidden width is fitted by bisection so that its parameter count
lands within $\pm5\%$ of $244{,}633$; the realised counts span
$244{,}317$ to $247{,}873$ ($-0.13\%$ to $+1.32\%$).
Table~\ref{tab:baseconf} records each arm's implementation and configuration.
Data, preprocessing, loss (Equation~\eqref{eq:loss}),
optimiser, epoch budget, early-stopping rule and evaluator are shared by all
seven arms. Two components differ: the encoder, and the readout---the six
baselines use the mean-and-last pooling head described above, while \model uses
the four-path readout of Section~\ref{sec:readout}. The two are not separated
by this table, so the reported comparison is between complete
encoder-plus-readout designs.

\begin{table}[t]
\centering
\small
\caption{Implementation and configuration of each arm. Baseline widths are set
by parameter count; all six use dropout $0.12$ and a shared mean-and-last
pooling head followed by a $64$-unit MLP.}
\label{tab:baseconf}
\begin{tabular}{lp{4.1cm}p{5.3cm}r}
\toprule
Arm & Implementation & Fixed configuration & Params \\
\midrule
MLP & \texttt{torch.nn} linear stack & two hidden layers ($h$, $h/2$),
  LayerNorm, GELU & 246{,}984 \\
LSTM & \texttt{nn.LSTM} & $2$ layers, inter-layer dropout & 245{,}381 \\
GRU & \texttt{nn.GRU} & $2$ layers, inter-layer dropout & 244{,}317 \\
TCN & ours, after \citet{bai2018tcn} & $4$ residual blocks, kernel $3$,
  dilations $1,2,4,8$, GroupNorm, causal left-padding & 245{,}201 \\
Transformer & \texttt{nn.TransformerEncoder} & $2$ pre-norm layers, $4$ heads,
  feed-forward $2\times$, learned positional embedding & 247{,}873 \\
Mamba (S6) & the selective-scan block implementation of \model's own backbone &
  $4$ blocks, single decay initialisation, linear stem & 245{,}413 \\
\model & frozen implementation of Section~\ref{sec:model} & $d=96$, $L=4$,
  $\mathcal{S}=\{3,12,48\}$ (Table~\ref{tab:params}) & 244{,}633 \\
\bottomrule
\end{tabular}
\end{table}

The structural hyper-parameters in Table~\ref{tab:baseconf} were fixed before
the reported runs; only hidden width was fitted, using parameter count rather
than accuracy. The shared loss and optimiser settings were developed with
\model, so the comparison represents standard configurations under one shared
recipe rather than per-architecture tuning.

\subsection{Training setup}

All arms are trained with AdamW~\citep{loshchilov2019adamw} at learning rate
$3\times10^{-4}$, weight decay $1.2\times10^{-3}$, gradient clipping at $0.8$,
two warm-up epochs, an exponential moving average of the weights with decay
$0.9985$ used for validation and checkpointing, mixup with $\alpha=0.1$ applied
to $30\%$ of batches, at most $60$ epochs, and early stopping with patience
$10$. The selection criterion is not any single metric but the equal-weight
combination of the percentile ranks of all four metrics of
Section~\ref{sec:protocol} within the run's own epoch pool. Every arm is trained
with three seeds ($42,43,44$); reported uncertainties are sample standard
deviations across seeds. Training runs on a single NVIDIA A800; \model takes
$2.91$ minutes per epoch. Determinism is enforced for all arms except TCN, for
which the deterministic backward kernel for dilated convolution is roughly two
orders of magnitude slower; that arm fixes seeds but does not guarantee
bit-level reproducibility.

\subsection{Main results}

Table~\ref{tab:main} reports the test year. \model leads on all four metrics.
Its rank IC of $0.0728$ is $14.9\%$ above the strongest baseline (GRU,
$0.0634$), its signal long-short Sharpe of $12.85$ is $17.7\%$ above the best baseline
Sharpe ($10.92$, GRU), and its stress IC\_IR of $1.030$ is $13.4\%$ above
GRU's $0.909$.

The seed-level test pairs the three seeds of each arm and gives $p\le0.014$ on
rank IC against every baseline, although inference over optimisation
randomness is necessarily coarse with three seeds. Separately, pairing the
$241$ daily cross-sectional ICs on the same days and cross-sections yields
Newey--West statistics (Bartlett, lag $5$) between $4.40$ and $6.29$, with
$p<0.001$ against every baseline (Table~\ref{tab:paired}). Seed dispersion and
HAC standard errors quantify optimisation and temporal sampling uncertainty,
respectively.

Two comparisons in the table carry more information than the overall ordering.
The first is \textbf{Mamba}, which shares \model's selective state-space
backbone but replaces the stem with a linear map, uses a single decay
initialisation, and pools with mean-and-last: it reaches $0.0615$ against
$0.0728$. The $18.5\%$ gap measures the joint contribution of the bundled stem,
decay-initialisation and readout changes over the common state-space backbone.
The second is \textbf{MLP}, the only arm with no temporal inductive bias, which
trails every sequence model by a wide margin ($0.0485$). The spread among the
five sequence baselines ($0.0586$ to $0.0634$) is far narrower than the gap
between them and MLP, so on this task \emph{having} a temporal structure
matters more than which one is chosen. The degree to which \model's margin
comes from the stem and readout rather than from other design choices
remains to be isolated by ablation.

The IC\_IR differences from GRU, Mamba and LSTM are not significant across
three seeds ($p=0.38$, $0.44$ and $0.29$, respectively).

\begin{table}[t]
\centering
\small
\caption{Test year (2024, 241 trading days) after residualisation against eight
style factors. Values are mean $\pm$ sample standard deviation over three seeds;
$p$ is the two-sided paired seed-level test against \model.}
\label{tab:main}
\begin{tabular}{lrrrrr}
\toprule
Model & Params & Rank IC & IC\_IR & Signal LS & Stress \\
      &        &         &        & Sharpe    & IC\_IR \\
\midrule
\textbf{\model (ours)} & 244{,}633 & $\mathbf{0.0728}_{\pm 0.0020}$ & $\mathbf{1.128}_{\pm 0.113}$ & $\mathbf{12.85}_{\pm 0.62}$ & $\mathbf{1.030}_{\pm 0.047}$ \\
\midrule
MLP         & 246{,}984 & $0.0485_{\pm 0.0009}$ & $0.788_{\pm 0.043}$ & $9.13_{\pm 0.18}$ & $0.586_{\pm 0.039}$ \\
LSTM        & 245{,}381 & $0.0604_{\pm 0.0006}$ & $1.026_{\pm 0.011}$ & $10.39_{\pm 0.07}$ & $0.836_{\pm 0.035}$ \\
GRU         & 244{,}317 & $0.0634_{\pm 0.0004}$ & $1.059_{\pm 0.005}$ & $10.92_{\pm 0.22}$ & $0.909_{\pm 0.016}$ \\
TCN         & 245{,}201 & $0.0593_{\pm 0.0005}$ & $0.924_{\pm 0.031}$ & $9.64_{\pm 0.23}$ & $0.768_{\pm 0.027}$ \\
Transformer & 247{,}873 & $0.0586_{\pm 0.0004}$ & $0.873_{\pm 0.014}$ & $10.21_{\pm 0.37}$ & $0.763_{\pm 0.027}$ \\
Mamba (S6)  & 245{,}413 & $0.0615_{\pm 0.0003}$ & $1.043_{\pm 0.042}$ & $10.76_{\pm 0.30}$ & $0.838_{\pm 0.030}$ \\
\midrule
\multicolumn{2}{l}{\emph{$p$ vs.\ \model}} & $\le0.014$ & $0.03$--$0.44$ & $\le0.050$ & $\le0.049$ \\
\bottomrule
\end{tabular}
\end{table}

\begin{table}[t]
\centering
\small
\caption{Day-level paired comparison on the test year ($241$ trading days,
residualised scores, seeds pooled). $\Delta$IC is the mean same-day difference
between \model and each baseline; standard errors are Newey--West (Bartlett,
lag $5$). The right block reports seed dispersion and temporal HAC uncertainty.}
\label{tab:paired}
\begin{tabular}{lrrrcrrr}
\toprule
& \multicolumn{3}{c}{paired against \model} & & \multicolumn{3}{c}{uncertainty in annual IC} \\
\cmidrule(lr){2-4}\cmidrule(lr){6-8}
Model & $\Delta$IC & $t$ & $p$ & & seed SD & HAC SE & ratio \\
\midrule
\model & --- & --- & --- & & $0.0020$ & $0.0042$ & $2.1\times$ \\
\midrule
GRU & $+0.00946$ & $5.54$ & $<0.001$ & & $0.0004$ & $0.0037$ & $9.3\times$ \\
Mamba & $+0.01139$ & $4.40$ & $<0.001$ & & $0.0003$ & $0.0036$ & $13.3\times$ \\
LSTM & $+0.01246$ & $5.23$ & $<0.001$ & & $0.0006$ & $0.0035$ & $5.8\times$ \\
TCN & $+0.01350$ & $5.92$ & $<0.001$ & & $0.0005$ & $0.0039$ & $7.4\times$ \\
Transformer & $+0.01421$ & $5.73$ & $<0.001$ & & $0.0004$ & $0.0040$ & $9.5\times$ \\
MLP & $+0.02433$ & $6.29$ & $<0.001$ & & $0.0009$ & $0.0038$ & $4.2\times$ \\
\bottomrule
\end{tabular}
\end{table}

\subsection{Style exposure of model scores}
\label{sec:r2}

For \model, residualisation lowers mean daily rank IC from $0.0893$ to $0.0728$
and its standard deviation from $0.0973$ to $0.0649$, raising IC\_IR from
$0.919$ to $1.128$. The same direction holds for every arm
(Appendix~\ref{app:survival}). We measure style exposure directly with the
per-day coefficient of determination of the residualising regression,
\begin{equation}
R^2_{\text{style}}(t) \;=\; 1 - \frac{\operatorname{Var}_t(\text{residual})}
{\operatorname{Var}_t(\text{standardised score})}
\label{eq:r2style}
\end{equation}
which is the fraction of that day's cross-sectional score variance explained
by the eight controls. Positive controls and a synthetic loading sweep validate
this measure and show why the ratio of residualised to raw IC is unsuitable for
the same purpose (Appendix~\ref{app:resid}).
\begin{table}[t]
\centering
\caption{Share of score variance explained by the eight styles in the daily
cross-sectional regression, averaged over test-year trading days and three
seeds. Lower values indicate less style-explained score variance.}
\label{tab:r2}
\begin{tabular}{lrr}
\toprule
Model & $R^2_{\text{style}}$ & $\pm$seed \\
\midrule
MLP                        & \textbf{0.1574} & 0.0221 \\
\textbf{\model (ours)}     & 0.1721 & 0.0078 \\
LSTM                       & 0.1797 & 0.0048 \\
GRU                        & 0.1842 & 0.0107 \\
TCN                        & 0.1874 & 0.0120 \\
Mamba (S6)                 & 0.2133 & 0.0192 \\
Transformer                & 0.2200 & 0.0085 \\
\bottomrule
\end{tabular}
\end{table}

A low $R^2_{\text{style}}$ must be interpreted jointly with predictive
performance. MLP has the lowest value ($0.157$) but is also the weakest
predictor, with a residualised rank IC of $0.049$. By contrast, \model combines
the highest residualised rank IC ($0.073$) with a lower
$R^2_{\text{style}}$ ($0.172$) than all five sequence baselines
($0.180$--$0.220$). Its predictive advantage is therefore not explained by
greater exposure to the eight controls. This conclusion is limited to the
price--volume factors considered here; omitted industry, valuation,
profitability, growth and leverage exposures may explain additional score
variation (Section~\ref{sec:limits}).

\subsection{Where the predictive power sits in the day}
\label{sec:exec}

The close-to-close label contains an overnight segment that precedes the score.
To quantify its contribution, we split the label into the overnight gap and the
following session,
\begin{equation}
\underbrace{\frac{C^{\mathrm{adj}}_{i,t+1}}{C^{\mathrm{adj}}_{i,t}}}_{1+y_{i,t}}
\;=\;
\underbrace{\frac{O^{\mathrm{adj}}_{i,t+1}}{C^{\mathrm{adj}}_{i,t}}}_{\text{not capturable}}
\;\cdot\;
\underbrace{\frac{C_{i,t+1}}{O_{i,t+1}}}_{\text{earliest executable}} ,
\label{eq:split}
\end{equation}
and rescore the stored predictions against each factor. The residualisation
pipeline of Section~\ref{sec:protocol} is applied identically in each case. The
open of day $t+1$ is taken from that
day's first five-minute bar, and the intraday factor needs no adjustment
because the adjustment factor is constant within a trading day
(Section~\ref{sec:data}), so it cancels from any same-session price ratio. The
split is available for
$100.0\%$ of test-year samples ($239{,}591$ observations over $241$ days).

\begin{table}[t]
\centering
\small
\caption{Stored scores under three return conventions (test year 2024, after
style residualisation, three seeds). Panel~A decomposes \model's label;
Panel~B compares all architectures over the executable session. $t$ is a
Newey--West statistic (Bartlett, lag $5$) on the daily long-short spread.}
\label{tab:exec}
\begin{tabular}{lrrrrr}
\toprule
& Rank IC & IC\_IR & \multicolumn{2}{c}{LS spread (bp/day)} & Signal LS \\
\cmidrule(lr){4-5}
& & & mean & $t$ & Sharpe \\
\midrule
\multicolumn{6}{l}{\textit{Panel A: \model, one score, three targets}}\\
\quad $C^{\mathrm{adj}}_{t}\rightarrow C^{\mathrm{adj}}_{t+1}$ \quad(label) & $0.0728$ & $1.128$ & $+73.9$ & $10.97$ & $12.85$ \\
\quad $C^{\mathrm{adj}}_{t}\rightarrow O^{\mathrm{adj}}_{t+1}$ \quad(overnight) & $0.1275$ & $2.431$ & $+75.6$ & $14.26$ & $24.13$ \\
\quad $O_{t+1}\rightarrow C_{t+1}$ \quad(executable) & $0.0215$ & $0.358$ & $-1.6$ & $-0.37$ & $-0.32$ \\
\midrule
\multicolumn{6}{l}{\textit{Panel B: all arms under $O_{t+1}\rightarrow C_{t+1}$}}\\
\quad \model & $0.0215$ & $0.358$ & $-1.6$ & $-0.37$ & $-0.32$ \\
\quad MLP & $0.0059$ & $0.100$ & $-11.8$ & $-2.99$ & $-2.67$ \\
\quad LSTM & $0.0149$ & $0.260$ & $-6.1$ & $-1.52$ & $-1.38$ \\
\quad GRU & $0.0162$ & $0.282$ & $-4.5$ & $-1.06$ & $-0.98$ \\
\quad TCN & $0.0132$ & $0.217$ & $-8.3$ & $-1.94$ & $-1.67$ \\
\quad Transformer & $0.0111$ & $0.172$ & $-10.5$ & $-2.75$ & $-2.38$ \\
\quad Mamba & $0.0157$ & $0.272$ & $-7.6$ & $-2.11$ & $-1.75$ \\
\bottomrule
\end{tabular}
\end{table}

Two conclusions follow, and they point in opposite directions.

\paragraph{The reported Sharpe ratio is concentrated overnight.} For all seven
architectures, the decile spread earned over the overnight gap
alone is \emph{larger} than the spread over the full close-to-close label
($+75.6$ against $+73.9$~bp per day for \model), and the spread over the
executable session is negative. Over that session, \model's top decile returns
$+6.0$~bp per day and its bottom decile $+7.5$~bp---both positive, with the
\emph{short} leg the stronger of the two. Stocks that gap down at the open
rebound during the session more than stocks that gap up continue, so the decile
spread reverses even though the cross-sectional ordering does not: the rank IC
over the executable session remains positive at $0.0215$.

\paragraph{Point-estimate ordering under the executable target.} Under the executable session the ordering of the seven arms is
unchanged and \model is still first. Its residualised rank IC of $0.0215$
exceeds the best baseline's by twice the relative margin it enjoys under the
close-to-close label---$+32.7\%$ over the strongest baseline (GRU, $0.0162$)
against $+14.9\%$ under the label---and its long-short spread of $-1.6$~bp per
day is the smallest in magnitude of the seven, with a Newey--West statistic of
$-0.37$ that cannot be separated from zero, whereas the weakest arms are
significantly negative (MLP $t=-2.99$, Transformer $t=-2.75$). The point-estimate
ordering is preserved under the executable target, although none of the
resulting long--short spreads supports a profitable strategy interpretation.

\subsection{Limitations}
\label{sec:limits}

\paragraph{One market, one horizon, one frequency.} All evidence comes from a
single national equity market, a one-day prediction horizon and five-minute
bars. The stem's scale set $\{3,12,48\}$ is expressed in bars and its largest
member coincides with one trading session; whether the architecture transfers to
markets with different session structures is untested.

\paragraph{The style set is a proxy.} The eight factors of
Section~\ref{sec:protocol} are computed from price and volume alone. They
contain no valuation, profitability, growth or leverage factor, and no industry
dummies. Residualising against them therefore removes less than a full
commercial risk model would, and the residualised metrics below should be read
as upper bounds on what would survive a stricter control.


\paragraph{Shared rather than per-arm tuning.} The loss and optimiser settings
were developed with \model and then applied to all architectures. Dedicated
per-arm tuning could narrow the reported gaps (Section~\ref{sec:baselines}).

\paragraph{Execution constraints.} Section~\ref{sec:exec} shows that the
close-to-close long--short spread is not executable from the signal timestamp.
The reported diagnostics also charge no transaction costs, slippage or
market impact, assume unconstrained daily rebalancing, ignore the daily price
limits that bind exactly on the days the signal is strongest, and assume
short availability that much of this universe lacks. A deployable strategy
would require an executable return convention, costs and borrowing constraints.

\section{Conclusion}

\model maps raw five-minute bars to next-day cross-sectional rankings with
244{,}633 parameters and no hand-built input features. Its multi-scale
zero-sum-initialised stem and staggered selective state-space backbone lead six
parameter-matched alternatives on all four metrics of a held-out year. For
rank IC, every day-level paired comparison has $p<0.001$, and among the five
sequence baselines \model has the lowest style-explained score variance. The
Mamba comparison attributes an $18.5\%$ rank-IC gap to the bundled design
changes; component-level contributions require ablation.

The close-to-close target includes an overnight move that precedes the score.
From the first executable price, the architecture ordering is preserved and
\model's relative rank-IC margin widens, but its long--short spread is
indistinguishable from zero. The results therefore support a prediction-model
comparison rather than a deployable trading strategy, and the style analysis
is limited to the eight price--volume controls used here.

\bibliographystyle{plainnat}
\bibliography{refs}

\clearpage
\appendix

\section{Corporate actions and the adjustment convention}
\label{app:adjust}

Inputs are raw exchange quantities; adjustment is applied only to the target.
Let $q_{i,e}$ denote the multiplicative backward-adjustment contribution of a
corporate action for stock $i$ with ex-date $e$ (including split and dividend
terms supplied by the data vendor). Define
\begin{equation}
 A_{i,d}=\prod_{e\le d}q_{i,e},\qquad
 C^{\mathrm{adj}}_{i,d}=A_{i,d}C_{i,d}.
\end{equation}
The product is over ex-dates on or before $d$ and is never revised by a later
action, so
\begin{equation}
 y_{i,t}=\frac{A_{i,t+1}C_{i,t+1}}{A_{i,t}C_{i,t}}-1.
\end{equation}
No $A_{i,d}$ appears among the $F=25$ input fields: prices, volumes and
order-book quantities enter exactly as printed. An ex-date inside the input
window therefore remains a raw mechanical jump for the network to observe,
while an action first entering through $A_{i,t+1}$ affects only the next-day
total-return label and cannot leak into $X_{i,t}$.

\section{Stress-day detector: full specification}
\label{app:stress}

Let $d_1<\cdots<d_M$ be the ordered evaluation dates and let $r_j$ be the
equal-weight market return for day $d_j$, computed from the same raw next-day
returns used for the long--short diagnostic. With $w=10$, form
\begin{align}
 c_j &= \sum_{k=j-w+1}^{j} r_k, &
 v_j &= \operatorname{sd}(r_{j-w+1:j}),
\end{align}
using the available suffix when at least $\lceil w/2\rceil=5$ finite returns
are present. For any series $x$, standardise over its finite entries:
\begin{equation}
 Z(x_j)=\frac{x_j-\operatorname{mean}_{k\in\mathcal F}x_k}
 {\max(\operatorname{sd}_{k\in\mathcal F}x_k,1)},
 \qquad \mathcal F=\{k:x_k\text{ is finite}\}.
\end{equation}
The stress strength and threshold are
\begin{equation}
 h_j=\max\{Z(|c_j|),Z(v_j)\},\qquad
 \tau=\operatorname{quantile}_{0.80}\{h_j:h_j\text{ finite}\}.
\end{equation}
A day is hot iff $h_j\ge\tau$; ties at the threshold are included and
non-finite strengths are cold. Hot positions are split into maximal runs.
Two runs are merged when the number of intervening cold positions is at most
$g=3$ (equivalently, the next start minus the previous end is at most $g+1$).
Only merged segments of length at least $m=5$ days are retained. This order
(merge, then length filter) and the position-based rule apply even when
missing days occur; missing days are never silently treated as hot.

The stress mask is the union of retained segments. Daily cross-sectional ICs
inside this mask are concatenated, and Stress IC\_IR is their mean divided by
their sample standard deviation; it is reported only when at least 20 finite
ICs remain. The detector uses whole-window means, standard deviations and the
empirical percentile, so it is intentionally non-causal; it is computed from
returns only, identically for every arm, and never supplied to the model. With
$(w,q,g,m)=(10,0.80,3,5)$, the 2024 test window yields 55 days in four
retained segments.

\section{Validation and interpretation of style residualisation}
\label{app:resid}

\subsection{Positive controls}
\label{app:controls}

We pass two signals lying in the span of the controls through exactly the
pipeline of Section~\ref{sec:protocol}: \textsc{rev5}$=-$\textsc{strev} and
\textsc{mom}=\textsc{momentum}. The symmetric protocol clips scores and
regressors at mean $\pm3\sigma$ before the cross-sectional OLS residual.
\begin{table}[t]
\centering
\caption{Positive controls on the 2024 test year ($241$ days). Residual IC is
the daily mean after the indicated clipping convention; SE is its sample
standard error across days; Survival uses the unmarked residual IC divided by raw IC. Symmetric clipping is the main protocol.}
\label{tab:naive}
\small
\begin{tabular}{@{}lrrrrrr@{}}
\toprule
Signal & Raw IC & Resid. IC & Resid. IC (sym.) & SE (sym.) & $R^2_{\rm style}$ & Survival \\
\midrule
\textsc{rev5} $=-$\textsc{strev} & $+0.00653$ & $+0.01522$ & $-0.00085$ & $0.00210$ & $1.000$ & $+2.33$ \\
\textsc{mom} =\textsc{momentum} & $+0.00604$ & $+0.01257$ & $+0.00196$ & $0.00200$ & $1.000$ & $+2.08$ \\
\bottomrule
\end{tabular}
\end{table}

Both symmetric residual ICs are within one SE of zero and have
$R^2_{\rm style}=1.000$. Under asymmetric clipping, the same controls retain
$+0.0152$ and $+0.0126$; symmetric clipping is therefore necessary to prevent
tail-induced residual artefacts.

\subsection{Why the survival ratio is not identified}
\label{app:survival}

The ratio of residualised to raw rank IC (``survival'') is not a style-exposure
measure. Write a score as $s=\mu\alpha+\lambda\phi+\nu\varepsilon$, with
standardised return $\alpha$, style $\phi$, independent noise
$\varepsilon$, and fixed $\mu=0.11$. Sweeping only $\lambda$ on the test
panel gives Table~\ref{tab:identif}: residualisation removes the style term,
but survival rises when loading on \textsc{strev} (own IC $-0.006$) and falls
when loading on $-\textsc{resvol}$ (own IC $-0.060$). The same alpha content
therefore produces opposite survival readings, whereas $R^2_{\mathrm{style}}$
changes monotonically with $|\lambda|$.

\begin{table}[t]
\centering
\caption{Controlled loading sweep with fixed true alpha content
($\mu=0.11$). Survival is non-monotone in the loaded style; style $R^2$ is
monotone.}
\label{tab:identif}
\begin{tabular}{lrrrrrr}
\toprule
& & \multicolumn{2}{c}{loading on \textsc{strev}} & & \multicolumn{2}{c}{loading on $-\textsc{resvol}$} \\
\cmidrule(lr){3-4}\cmidrule(lr){6-7}
$\lambda$ & & Survival & $R^2_{\mathrm{style}}$ & & Survival & $R^2_{\mathrm{style}}$ \\
\midrule
$0.00$ & & $0.824$ & $0.011$ & & $0.824$ & $0.011$ \\
$0.15$ & & $0.852$ & $0.034$ & & $0.763$ & $0.035$ \\
$0.30$ & & $0.908$ & $0.099$ & & $0.733$ & $0.102$ \\
$0.45$ & & $0.990$ & $0.205$ & & $0.732$ & $0.212$ \\
$0.60$ & & $1.113$ & $0.348$ & & $0.767$ & $0.363$ \\
\bottomrule
\end{tabular}
\end{table}

\begin{table}[t]
\centering
\caption{Raw and residualised rank IC over the test year (three seeds). The
survival ratio is shown for context, not used as a style-dependence metric.}
\label{tab:survival}
\begin{tabular}{lrrrr}
\toprule
Model & Raw IC & Residualised IC & Survival & $\Delta$ IC\_IR \\
\midrule
\textbf{\model (ours)} & \textbf{0.08931} & \textbf{0.07284} & $0.816$ & $+22.8\%$ \\
GRU         & 0.08244 & 0.06337 & 0.769 & $+24.6\%$ \\
Transformer & 0.07650 & 0.05863 & 0.766 & $+31.7\%$ \\
LSTM        & 0.07947 & 0.06038 & 0.760 & $+32.3\%$ \\
TCN         & 0.07823 & 0.05933 & 0.758 & $+26.5\%$ \\
Mamba (S6)  & 0.08324 & 0.06145 & 0.738 & $+39.3\%$ \\
MLP         & 0.06917 & 0.04851 & 0.701 & $+24.5\%$ \\
\bottomrule
\end{tabular}
\end{table}

\end{document}